# Haderach Principle:
# symbiosis of scientific formalism and informal perspectives

Eldar Knar[1]

Tengrion, Astana, Republic of Kazakhstan
eldarknar@gmail.com
https://orcid.org/0000-0002-7490-8375

**Abstract**

Current research funding systems are subject to structural imbalances, where formal criteria such as the H-index and the number of publications dominate over the potential and actual scientific prospects of researchers. This leads to the suppression of potential breakthrough research directions and limited access to grants for stochastic (innovative) researchers. In this paper, we propose a dynamic agent-based model of research grant redistribution that takes into account the adaptive mechanism of funding redistribution on the basis of the quality of stochastic research.

The simulation was conducted on a sample of 21,534 Kazakhstani researchers with 30 iterations, during which the growth of the formal features of stochastic scientists was analyzed under different grant distribution scenarios. A grant redistribution parameter $\lambda$ was introduced, which controls adaptive funding. The results showed that at $\lambda=0.15$, stochastic scientists begin to catch up with formal scientists in terms of productivity without destabilizing the scientific system.

On the basis of the data obtained, a principle called the Haderach principle was proposed. It consists of a dynamic balance between formal (stable) and stochastic (informal) science. The developed approach can be used to optimize grant systems, allowing the elimination of barriers to new scientific directions and potential achievements without losing the stability of traditional schools.

New concepts and terms of scientific vocabulary are introduced: the Haderach principle, excluded science, supplemented science, grant monopoly, and so on.

Article structured V new IRPAS (Induction, Related Works, Processing, Analysis, Synthesis) notations.

**Keywords:** scientific grants, grant system, scientific funding**,** grant redistribution, stochastic science, agent-based modeling, resource allocation model, Haderach principle

**Declarations and Statements:**
No conflicts of interest
This work was not funded
No competing or financial interests
All the data used in the work are in the public domain.
Generative AI (LLM or other) was not used in writing the article.
Ethics committee approval is not needed (without human or animal participation).

---

[1] Fellow of the Royal Asiatic Society of Great Britain and Ireland

1. **Induction**

Even nonhuman primates reject the idea of an unfair and uneven distribution of incentives for work and remuneration for work [2]. Human scientists, especially those very rarely agree with unfairness, inequality, elitism and credentialism (selection only by formal indicators) in systems of financial incentives for scientific research. Moreover, all these shortcomings and flaws inevitably lead to an ineffective and unproductive scientific system.

The overwhelming majority of scientific grant systems are focused on deterministic criteria. For example, formal indicators (degree, title, publications, position) or the Hirsch index. This is what we call the principle of verified success. This is quite understandable and justified. Grant systems, especially if they are formed at the expense of the state budget, do not like risks. They are focused mainly on guaranteed results, which can be obtained with the highest probability from "verified" scientists with a fixed background. In turn, scientists who do not have formal indicators are left out of work, even if they propose quite original and progressive ideas. The de facto ban on grant funding through the requirement for formal indicators does not allow them to materialize projects and solutions that could become triggers for higher-level science. Formal and informal scientists can make scientific breakthroughs. However, the institutional conditions for providing grants prevent the participation of informal scientists in scientific progress [3]. From an official point of view, this cutoff is justified: they have not "earned" it because they do not meet the qualification requirements. Therefore, in terms of scientific progress, they rely entirely on "proven" and distinguished scientists. This is logical but ineffective. This is especially the case if the expectations do not match the possibilities in cases of the "average science trap".

Therefore, such a scheme works very well for preserving the current status quo of the scientific system. However, grant extractiveness does not work very well when real and impressive progress in national science is needed. This dilemma is especially relevant if we consider the status quo of average, below average and unremarkable science in terms of indicators and parameters. It appears to be you want to develop science, but the fear of losing what has already been achieved and the phobia of the unknown often prevail. More often than not, the bureaucracy would prefer to finance a distinguished, respected and proven professor of a university or institute with many years of awards and insignia rather than some incomprehensible and unknown young

---

[2] Brosnan , S. F., & de Waal, F. B. M. (2003). Monkeys reject unequal pay. Nature, 425(6955), 297–299. https://doi.org/10.1038/NATURE01963

[3] Laudel , G., & Glaser , J. (2014). Beyond breakthrough research: Epistemic properties of research and their consequences for research funding. Research Policy, 43(7), 1204–1216. https://doi.org/10.1016/J.RESPOL.2014.02.006

"Perelman " or aspiring " Galois" almost from the street. Even if the professor has not created anything outstanding or, as they like to say, breakthrough. This, in fact, explains why scientific systems with higher administrative and academic bureaucratization have lower scientific efficiency [4].

This is what can be called credentialism in science or what we call the "*excluded science effect*". This is a completely natural phenomenon for conservative and bureaucratized scientific systems. The main source of expenditure on science is the state budget. Private funding is absent or minimized. This does not mean that informal scientists are better than formal scientists. We are talking about the main thing—egalitarianism [5] in the context of grant opportunities. Which some have, others do not.

Equal opportunities, not on the basis of social status but on the basis of probable prospects or a seemingly talented project, constitute a direct and obvious path to the development of "good" and "excellent" sciences.

Thus, one of the key consequences of this system is the limitation of opportunities for stochastic scientists who do not have access to funding, which leads to stagnation of their scientific development. As a result, their H indices remain low, which prevents them from meeting the formal criteria for receiving grants. Thus, a vicious circle arises in which science with a high formal rating suppresses potential high science, leaving promising research on the periphery. That is, "formal" scientists have a high H-index (20--50) or official indicators and statuses (head, doctor, professor, etc.) since they deserve stable funding. In addition, "stochastic" scientists have low H indices (1--5) since they do not have access to grants and cannot actively publish. Here, we are talking not about superiority but about circumstances. Moreover, the fetishization of the H-index leads to a distortion of the assessment of the real scientific potential of researchers [6].

It seems that scientific research in the context of grant funding should be divided into two categories:

- Formal (deterministic) scientific projects based on the principles of credentialism. When grants are awarded on the basis of achievements, formal indicators and a system of official gradations,

- Informal (stochastic) scientific projects, when grants are awarded solely on the basis of their scientific intensity, a priori scientific extraordinary nature and potential "breakthroughs" without reference to the scientific status of the scientist.

---

[4] Coccia , M. (2009). Bureaucratization in public research institutions. Minerva , 47(1), 31–50. https ://doi . org/10.1007/s 11024-008-9113 -z

[5] Vaesen , K. K., & Katzav , J. J. (2017). How much would each researcher receive if competitive government research funding were distributed equally among researchers. PLOS ONE, 12(9), 1–11. https://doi.org/10.1371/JOURNAL.PONE.0183967

[6] Knar, E. (2024). Recursive index for assessing value added of individual scientific publications. arXiv preprint arXiv:2404.04276.

Notably, the credential system of grant funding completely dominates, for example, in the Kazakhstani grant system. There is no spontaneity or stochasticity here. As a result, the scientific system does not emerge from the state of the "average science trap" [7]. Despite all the declarations and plans. To change for the better, this formal dominance should be diluted with the right to informal grant recognition.

In accordance with this, we postulate and present a dynamic grant distribution model. This is based on the adaptive redistribution of funding between "formal" and "stochastic" sciences. In particular, we model a system of scientific funding in which stochastic scientists receive limited, but quite tangible, access to grants. We introduced a dynamic redistribution parameter that formats how funding changes depending on the quality of stochastic science. We model how the H-index (including formal indicators and statuses) of "stochastic" scientists changes when accessing funding and compare it with that of formal scientists and identify the optimal value at which science remains systemically stable, but stochastic scientists have the opportunity to catch up with formal scientists in terms of productivity and efficiency.

Accordingly, we have developed an agent-based model for the distribution of scientific grants, which includes the following attributes and definitions:

- "Formal" scientists, that is, researchers with a scientific background who receive the bulk of the funding.
- " informal" scientists, that is, researchers with little or no background , potentially capable of scientific breakthroughs, but without the resources to realize their potential,
- variable $\lambda$, which formats the redistribution of funding in favor of stochastic science depending on its quality,
- Multiple grant redistribution scenarios to assess the impact of different grant strategies.

In general, this study allows us to understand how modernization of the grant research funding system can eliminate artificial barriers between different categories of scientists and ensure adequate support for promising areas of science or extraordinary and promising research projects.

Thus, this study focuses on the conceptualization of an inclusive and authentic grant policy that allows traditional scientific positions and nonstandard directions to be balanced.

Thus, in this paper, a grant system based on formal features with a grant subsystem based exclusively on scientific content, effective scientific processing and the scientific intensity of scientific projects is proposed. In the scientific system, this creates the effect of "augmented science", which significantly increases the efficiency and effectiveness of integrated national science. For example, a standard grant can be received only by a scientist with an authentic set of qualification requirements (formal

---

[7] Knar, E. (2024). Homeopathic Modernization and the Middle Science Trap: conceptual context of ergonomics, econometrics and logic of some national scientific case. arXiv preprint arXiv:2411.15996

features). However, for example, a "publication grant", a grant for payment of APC in Q 1-Q 2 journals, can be received by any scientist. This does not require formal achievements, but an official notification to the editors for publication is sufficient. Considering the value in itself of the fact of publication in highly rated Q 1 and Q 2 journals. Moreover, a publication grant may not be issued directly to a scientist. However, it can be paid through a centralized system upon the provision of sufficient grounds.

This makes sense. The state, in some cases, does not need to issue grants for research (tens or even hundreds of millions of tenges) with subsequent publication effects. In addition, it will be enough for it to pay a very small amount for the publication itself in highly rated journals. The financial difference here is several orders of magnitude.

This is just one example out of dozens that a priori gives the effect of maximum optimization of the grant scientific system due to the symbiosis of formal (trendential) and informal (excluded, stochastic) sciences.

## 2. Related Works

The issue of correct scientific policy on expenditures on science and grant funding of scientific projects has the most direct and immediate relation to high-quality and effective science [8]. With an overly centralized system of funding science, the aforementioned danger of creating a "club of old friends" instead of a competitive environment in the "pursuit of grants" arises. Two key perspectives logically follow from this perspective: the "influence" perspective, which seeks the effects of specific management mechanisms, and the "influence on" perspective, which asks what factors contribute to the construction of the content of research [9].

Here, another circumstance arises that significantly increases the gap between the granted and the nongranted scientists. This is the principle of open science. Within the framework, the probability of publication increases with the payment of so-called editorial fees. The overwhelming majority of editorial offices are gradually switching to this model from a subscription model or even a hybrid model. As a result, the "club of old friends" significantly expands the "corridor of opportunities" for publication. Since they have a certain reserve of grant funds, payment for publications is provided. Thus, formal scientists with almost guaranteed access to grants, such as "the rich get even richer" [10], in the context of publication baggage as an argument for grant privileges.

---

[8] Meirmans, S., Butlin, R. K., Charmantier, A., Engelstädter, J., Groot, A. T., King, K. C., ... Neiman, M. (2019). Science policies: How should science funding be allocated? An evolutionary biologists' perspective. Journal of Evolutionary Biology, 32(8), 754–768. https://doi.org/10.1111/jeb.13497)

[9] Glaser, J., & Laudel, G. (2016). Governing Science: How Science Policy Shapes Research Content. Archives Europeennes De Sociologie, 57(1), 117–168. https://doi.org/10.1017/S0003975616000047

It is obvious that scientists who are engaged in new and promising areas should have greater preferences than scientists who are engaged in "proven" science with fewer prospects. New articles bring relatively greater benefits to science [11]. However, the problem here is that "new" projects and ideas most often receive lower ratings from experts and grant assessors [12]. Accordingly, there are fewer chances of receiving scientific grants, sometimes because of the effect of "intellectual distance" [13]. As a result, these scientists are falling further behind the "scientific schedule". In addition, they begin to meet formal requirements and criteria less and less. Move into the group of informal scientists with minimal chances of receiving scientific grants.

Accordingly, science funding strategies require a certain or even radical adaptation to certain specific conditions of the scientific system or the structure of the research system [14].

The issues of implementing the mechanism of effective distribution of finances for science remain relevant [15]. In addition, here the most diverse proposals, options, alternatives and schemes are being implemented [16].

There is, however, a strong scientific argument [17] that scientific efficiency is achieved to a greater extent by distributing grant resources among a larger number of scientists and scientific groups than by localizing them among a relatively small group of "elite" formalized scientists . In particular, funding strategies that target

---

diversity rather than "excellence" are probably more productive [18]than localized grant funding.

Therefore, the effectiveness and efficiency of scientific systems are determined by how inclusion and diversity are interpreted in grant programs. We do not mean social but rather intellectual inclusion.

### 3. Processing

In this paper, we accept the following *null hypothesis*:

*- the average values of the formal features of formal and stochastic scientists do not differ. That is, structurally, grant funding does not have a significant effect on the scientific productivity of scientists.*

And *an alternative one*:

*- the average values of the formal characteristics differ significantly. That is, there is a statistically significant effect of the grant funding structure.*

Despite their apparent obviousness and triviality, these hypotheses make some sense in the context of the problem under consideration. If the difference is statistically significant, then traditional grant funding, which is based on the principles of formalism and credentialism, clearly and definitely forms artificial barriers for stochastic science.

Hypothesis testing and statistical analysis of the criteria were interpreted through numerical modeling (including "running" of scenarios) in Python in the Jupyter environment. Notebook.

The initial state of the grant financing system is interpreted through the input parameters of the independent variables:

*Table 1.* Independent variables

| Variable | Designation | Description |
|---|---|---|
| lambda_values | $\lambda$ | Grant redistribution coefficient (0, 0.15, 0.3) |
| num_agents | $N$ | Number of scientists in the system (21,534) |
| num_cycles | $T$ | Number of simulation iterations (30) |
| W_total | $W$ | General grant funding fund (100 units) |
| W_formal_init | $W_F$ | Seed funding for formal science (100% of the grant program). |
| W_stochastic_init | $W_U$ | Seed funding for stochastic science (0% of grant program) |
| Q_stochastic | $Q$ | Quality of stochastic studies (0–1) |
| is_stochastic | $S$ | Scientist type (0 – formal, 1 – stochastic) |

---
[18]Fortin, J. M., & Currie, D. J. (2013). Big Science vs. Little Science: How Scientific Impact Scales with Funding. PLOS ONE, 8(6). https://doi.org/10.1371/JOURNAL.PONE.0065263

The key independent variable is the grant redistribution coefficient λ as a form factor of the evolutionary dynamics of the grant system. It controls whether grants are redistributed, and without it, the model is strictly deterministic.

As the number of scientists in the system, we took a value of 21,534. This number represents the number of Kazakhstani scientists as of 2023 according to the Bureau of National Statistics of the Republic of Kazakhstan [19]. We took the maximum sample. Not all scientists can make a significant contribution to science, but undoubtedly, every scientist, regardless of rank, has the probability of generating new knowledge. That is, every scientist (from a master's degree to an academician) can propose a hypothesis, idea or project of a high or extraordinary scientific level. If he is not a member of the club of formally selected scientists, then he has the right to receive a stochastic grant.

The dependent variables change during the modeling process and depend on the input parameters:

*Table 2.* Dependent variables

| Variable | Designation | Description |
| --- | --- | --- |
| W_formal | $W_F(t)$ | Funding formal science at iteration t |
| W_stochastic | $W_U(t)$ | Funding Stochastic Science at Iteration t |
| Delta_W | $\Delta W$ | Changes in funding for stochastic science |
| H_index | $H$ | Formal feature (usually this is the H-index) |
| avg_H_formal | $W_F$ | Average H of formal scientists |
| avg_H_stochastic | $W_U$ | Average H of stochastic scientists. |
| h_growth_rate | $\Delta H$ | Growth rate H of stochastic scientists |

The main criterion for the effectiveness of the model is the difference in the growth of H of stochastic scientists for different values of λ.

The control variables do not change during the modeling process, but they affect the dependent variables (Table 3). They eliminate side effects and ensure that the analysis is sufficiently correct.

Table 3. Control variables

| Variable | Designation | Description |
| --- | --- | --- |
| phi | $\Phi$ | The golden ratio (1.618) used for the staged distribution of grants |

---

[19] Key indicators of research and development work in the Republic of Kazakhstan. Series 19 Statistics of education, science and innovation, March 20, 2024, BNS ASPIR RK

| *seed* | - | Fixed value for random number generator (np.random .seed (42)) |
| *equal_var = False* | - | A condition in Welch's t test that allows for different distributions of variance to be taken into account |
| *num_bins* | - | Number of bins in the histogram H |

## 4. Analysis

### 4.1 *Big Picture*

Thus, a priori and from some of our own observations, we can state that in some conservative and static scientific systems (for example, in Kazakhstan), stable institutional inequality in the context of science dominates. Undoubtedly, scientific grants are among the most effective instruments for supporting scientific activity and research [20]. However, grant programs and funding of science in general (including the lack of inclusive basic funding in the salary part) are often uneven and fragmented. Some researchers are left without funds for scientific self-development and the development of a specific scientific direction. The pressure of financial deficit has led to the erosion of the middle class and disadvantaged scientists, while the scientific elite continue to concentrate resources on self-reproduction and maintaining the current scientific status quo. Scientists who constantly receive grants or even several grants simultaneously guarantee their inviolability regardless of the dynamics of efficiency and effectiveness. In fact, being in a state of consolidation of the material covered. Those scientists who do not receive grants or receive them episodically, sooner or later fall out of the scientific system. This rejection is not necessarily accompanied by forced emigration. Often, such scientists become the "work force" in other successful groups or projects. That is, they join the established "scientific traditions" and "scientific schools", counting on access to grants [21]. In general, in comparison with individual implementation, team implementation of science is becoming increasingly effective [22], but informal and individual implementations still make scientific breakthroughs. This circumstance must be taken into account.

Thus, this stable structural inequality becomes the main resistance factor, which directly or indirectly suppresses the prospects and progressiveness of the scientific system. The scientific closed hierarchy is constantly being improved and strengthened. In particular, new restrictive rules or definitions have been introduced (for example, the so-called "leading scientists "). If in developed scientific systems, the concept of "leading scientist" (and other similar terms) is unwritten and informal, then, for example, in Kazakhstan and Russia, this concept is quite formal and legislative.

---

[20] Azoulay , P., & Li, D. (2021). Scientific Grant Funding. SSRN Electronic Journal. https://doi.org/10.2139/ssrn.3563957

[21] Ebadi , A., & Schiffauerova , A. (2015). How to Receive More Funding for Your Research? Get Connected to the Right People! PLOS ONE, 10(7), 1–19. https://doi.org/10.1371/JOURNAL.PONE.0133061

[22] Wuchty , S., Jones, B. F., & Uzzi , B. (2007). The Increasing Dominance of Teams in Production of Knowledge. Science, 316(5827), 1036–1039. https://doi.org/10.1126/SCIENCE.1136099

In this scientific environment, they like to talk about the need to "raise the bar." This narrows the "opportunity corridor" for many other scientists, leaving themselves out of competition. This increases the likelihood that established scientific groups will have guaranteed scientific support. This problem is directly related to the old but still relevant "Matthew effect [23]." We believe that this problem is related not only to the classic "struggle for survival" in limited scientific systems but also to the phobia of erosion of the scientific image in the context of the Dunning–Kruger effect.

In general, such a situation leads to the degradation of the scientific system, the monopolization of resources, the erosion of development prospects, the imitation of vigorous scientific activity and, most importantly, the erosion of research motivational emergence [24]. The grant system becomes ineffective, and state funds for science go virtually nowhere.

These conservative and bureaucratic scientific systems are structurally analogous to Cantor sets [25], where at each iteration stage, new layers of scientists are excluded from the grant funding system. In addition, resources are even more concentrated in elite groups.

This happens, as we have already noted, due to new restrictive rules in competitive conditions or grant requirements. For example, in the next iteration, the need to "educate a doctoral student" is introduced into the grant conditions. Another group of scientists falls outside the grant Cantor set, since owing to objective circumstances, they cannot fulfill this condition. In addition, so on.

With respect to formalization, owing to the strengthening of requirements in the context of grant credentialism, the scientific system, from the point of view of grant financing, increasingly resembles a "scientific club of interest" or a "club of old friends" [26]. Cantor dynamics increasingly rarefies scientific space, leading to the loss of scientific potential and development prospects.

Thus, the scientific system is interpreted through conformism. When scientists stop generating new (i.e., risky) scientific projects and focus on proven "safe" topics with guaranteed publication and reporting. New scientific ideas and projects require time and risk, and the bureaucratic system usually does not finance projects that may not yield immediate or short-term results.

---

[23] Merton, R. K. (1968). The Matthew effect in science. The reward and communication systems of science are considered. Science, 159(3810), 56–63. https://doi.org/10.1126/SCIENCE.159.3810.56

[24] Knar, E. (2025). Optimal Salaries of Researchers with Motivational Emergence. arXiv preprint arXiv:2502.17271

[25] The Cantor set is formed by removing the central part of the segment at each iteration.

[26] Berezin , A. (1998). The perils of centralized research funding systems. Knowledge , Technology & Policy, 11(3), 5–26. https://doi.org/10.1007/S12130-998-1001-1

In general, the scientific system is being transformed into a "factory of average publications." In which "leading scientists" set trends, funding requires increasing formalization, and grants are awarded only on the basis of "proven success."

The proponents of trendionalism, i.e., the dominance of formal indicators, proceed from the dubious principle of "*funding people, not projects*" [27]. That is, if a scientist has good formal indicators but not a very good scientific project, he or she has a greater chance of success in a grant. Compared with a good project, a good project is proposed by a scientist with fewer scientific achievements and formal criteria. This also concerns the increase in the "career gap" between formal and informal scientists as a factor in grant preference. The career of a scientist is a function not only of scientific potential but also of the amount of funding [28]. In particular, funded articles demonstrate superlinear growth in citations, exceeding the growth observed in unfunded articles [29].

As a result, a situation of dominance of realized science over unrealized and potentially promising science arises in the scientific system. That is, formalization, credentialism and average realized science (based on Hirsch indices, affiliations, publications and formal requirements) suppress potentially highly unrealized science and potential scientific "breakthrough" ideas.

This is the main thesis and null hypothesis of this work.

Therefore, in "correct" science, the principle of inclusive equal opportunities for formal (proven) and informal (uncertain) scientists in grant systems with state dominance must be implemented to achieve resonance in the efficiency and effectiveness of modern and future scientific research.

We called this the Haderach Principle (Kwisatz Haderach (meaning "shortening the path"), named after the hero of the Dune Chronicles Frank Herbert, who had the foresight and ability to find optimal balances between the past and the future.

### 4.2. Models and imitation

Let us define the concept of a grant monopoly as a situation in which science funding is predominantly carried out through a grant system, basic funding is minimal, business is almost not involved in funding and patronage of scientific research, and science expenditures are exclusively or predominantly a state prerogative. A grant monopoly is characteristic of the overwhelming majority of post-

---

[27] Shaw, J. (2024). "Fund people, not projects": From narrative CVs to lotteries in science funding policy. Research Evaluation, 33. https://doi.org/10.1093/reseval/rvae035

[28] Goldfarb, B. (2008). The effect of government contracting on academic research: Does the source of funding affect scientific output? Research Policy, 37(1), 41–58. https://doi.org/10.1016/j.respol.2007.07.011

[29] Coccia, M., & Roshani, S. (2024). General laws of funding for scientific citations: how citations change in funded and unfunded research between basic and applied sciences. Journal of Data and Information Science. https://doi.org/10.2478/jdis-2024-0005

Soviet and conservative national scientific systems. In particular, a grant monopoly exists in the Kazakh scientific system.

In such systems, grants are strictly determined in the context of formal criteria, and the degree of determination depends on the paradox of scientific conservatism. This can be interpreted as follows:

$$P \sim \frac{H}{(1+R)}$$

where the probability of receiving a scientific grant $P$ is proportional to the formal criterion $H$ and inversely proportional to the bureaucratic resistance of the grant system $R$.

To modernize the system, we propose including a part of the scientific community without formal characteristics in scientific grant processing. The balance between formal and informal science in the context of grant finances can be a priori interpreted through the golden section as a hypothetically optimal ratio:

$$W = 0.618\,W + 0.382 W = W_F + W_U,$$

where $W$ is the total budget of the grant program, $W_F \approx 61.8\%$ - funding of formal science and $W_U \approx 38.2\%$ - funding of innovative ideas.

If stochastic science begins to produce significant scientific results in this case, the grant system automatically redistributes funds:

$$\Delta W_U = \lambda\,(Q+L)$$

where $\Delta W_U$ is the increment of grant funding for stochastic science, $\lambda$ is the coefficient of redistribution of grant funds, $Q$ is the quality of scientific results, and $L$ is the average level of scientific work.

If a grant funding management system follows this principle, it will naturally maintain a balance between reproducible science and potential scientific breakthroughs or additional scientific results without external regulation.

The recurrence relation of the distributed grant system is interpreted on the basis of the quality of stochastic science:

$$W_{U,n+1} = W_{U,n} + \lambda(Q_n - L_n)$$

$$W_{F,n+1} = W_n - W_{U,n+1}$$

On the basis of the results of numerical modeling, the dynamics of the grant system were obtained on the basis of the parameter λ (Table 4). Comprehensive summary table).

For λ=0, formal (deterministic) science $W_F$ receives 100% of grant funding, whereas stochastic (excluded) science $W_U$ receives almost 0%. Naturally, the constraint of the deficit of opportunities leads to a low value of formal indicators H (average value ≈ 2.9). This is the final state of the scientific system as a result of credentialism.

At λ=0.15, the redistribution of grants occurs adaptively. That is, formal science accounts for approximately 59%, and stochastic science accounts for approximately 41%. As a result, the average value of H for informal scientists increases sharply (≈ 25.9). However, for formal scientists, it remains stable (≈ 34.6). This result is obvious and trivial. However, the peculiarity here is that not only formal scientists but also informal scientists are involved in grant processing, who hypothetically and actually can be creators of scientific progress. However, they are limited in funds because of the pressure of credentialism.

At λ=0.3, the redistribution becomes aggressive. The informal value is approximately 45%, which leads to an even higher average value of H (strengthening of formal features, such as the H-index) (≈ 28.1). However, at the same time, naturally, the funding of formal science decreases to ≈ 54.8%. In this case, we do not mean a quantitative reduction in grant spending on formal science. After all, we a priori believe that the total volume of spending on science should grow permanently. Therefore, in this case, we consider only proportional relationships.

*Table 4.* Complex summary table

| λ | $W_F$ % | $W_U$ % | Δ W | meanH$_F$ | medianH$_F$ | $H_F$ | meanH$_U$ | medianH$_U$ | $H_U$ |
|---|---|---|---|---|---|---|---|---|---|
| 0 | 100 | 0 | 0 | 35.2 | 35 | 8.5 | 2.9 | 3 | 1.5 |
| 0.15 | 59.1 | 40.9 | 1.8 | 34.6 | 35 | 8.2 | 25.9 | 26 | 10.1 |
| 0.3 | 54.8 | 45.2 | 3.4 | 33.2 | 33 | 8 | 28.1 | 28 | 11.3 |

With a fixed λ=0.15, grants are distributed smoothly (Table 5). This allows informal scientists to increase grant activity and, accordingly, formal indicators due to stochastic (noncredential) granting (Graph 1).

*Table 5.* Evolutionary dynamics of grant distribution by iterations for a fixed value of λ.

| Cycle | λ | W_F % | W_U % | Δ W |
|---|---|---|---|---|
| 0 | 0.15 | 61.8 | 38.2 | 0.5 |
| 5 | 0.15 | 60.4 | 39.6 | 1.2 |
| 10 | 0.15 | 59.1 | 40.9 | 1.8 |
| 15 | 0.15 | 58.3 | 41.7 | 2 |
| 20 | 0.15 | 58 | 42 | 2.1 |
| 25 | 0.15 | 58.2 | 41.8 | 2 |
| 30 | 0.15 | 59.1 | 40.9 | 1.8 |

*Graph 1.* Dynamics of the stochastic science grant distribution for different values of λ

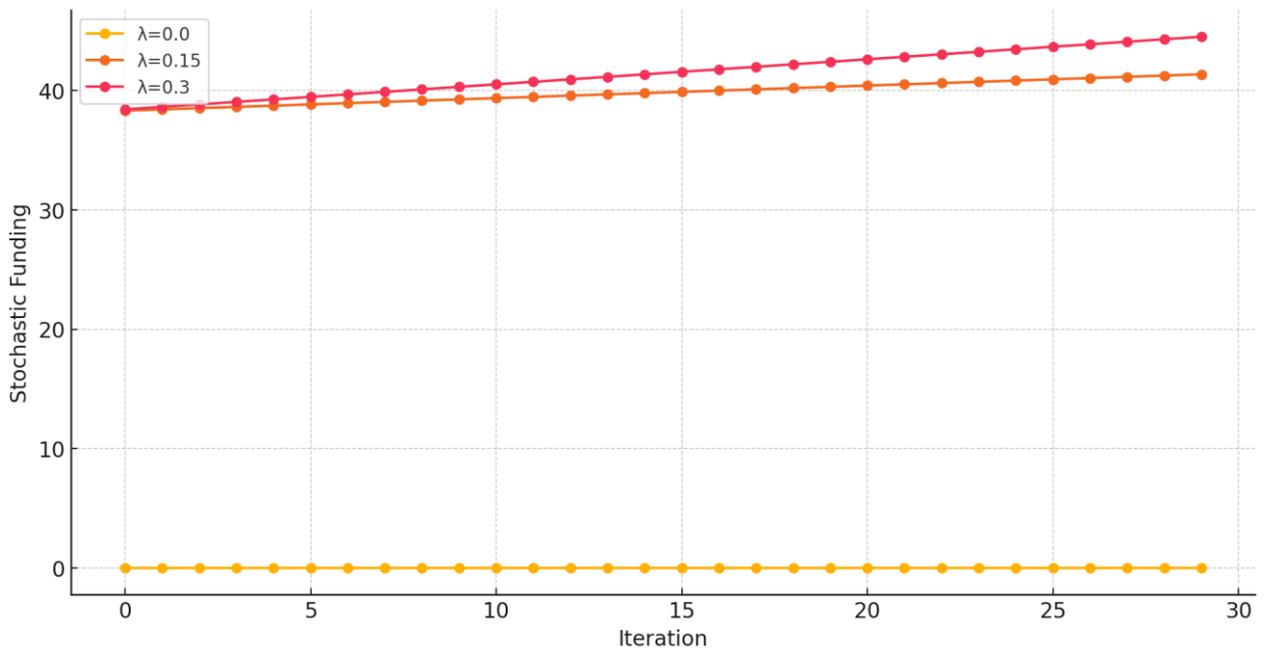

The graph shows a horizontal line at zero for λ= 0, a smoothly increasing line for λ=0.15 and more rapid growth for λ=0.3.

The formal features of H formal scientists remain virtually unchanged (Table 6). This corresponds to the situation in the scientific system that we previously called the "*average science trap*".

Table 6. Evolutionary dynamics of formal attributes

| Cycle | H F | H U |
|---|---|---|
| 0 | 35.4 | 2.9 |
| 10 | 35.1 | 10.2 |
| 20 | 34.5 | 18.5 |
| 20 | 34 | 25.9 |

The formal attributes of informal scientists increased dramatically from 2.9 to 25.9. That is, access to grant funding bypassing qualification requirements (formal attributes) provides the opportunity for "excluded " science to make its contribution to the overall national science. Of course, it is not as significant as formal science but is ideally approaching it in terms of formal achievements and recognition. Here, as they say, options are possible.

At λ=0.0, the stochastic group line is practically zero; at λ>0, grants are gradually redistributed in stochastic science (Graph 2).

*Figure 2.* Comparison of formal and stochastic science funding by iterations for

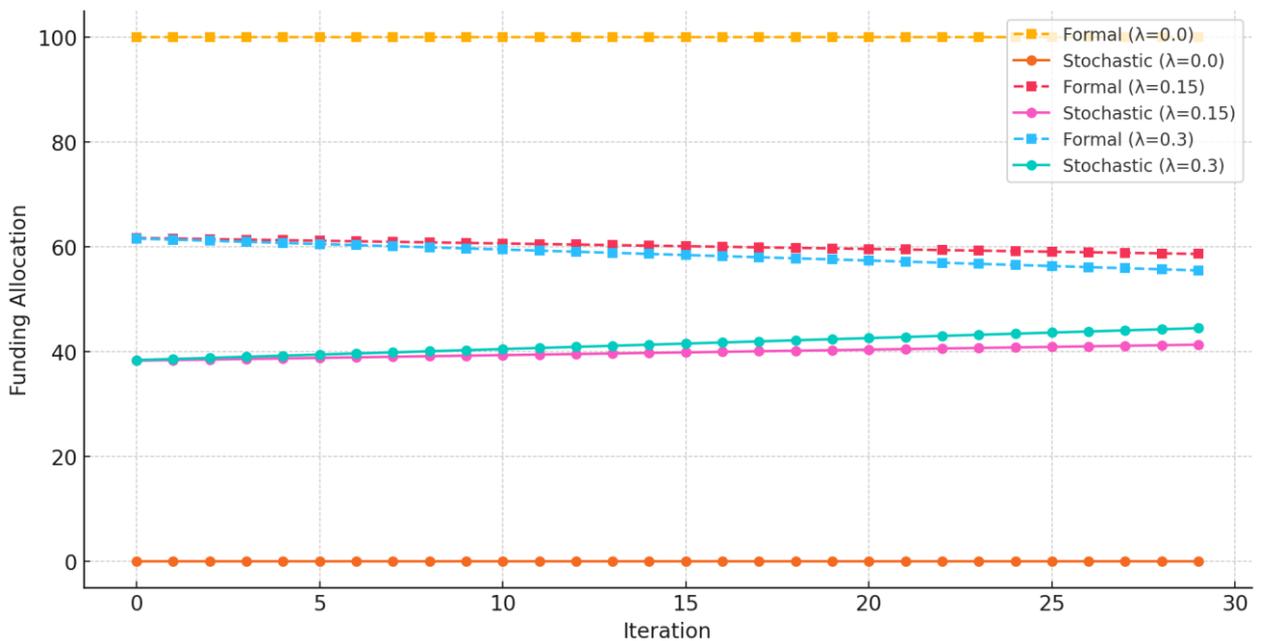

When formal features are redistributed for formal scientists, one line remains stable (formal scientists), whereas the line for stochastic scientists rapidly increases from low values to significantly higher values. As a result, the distribution of formal features of stochastic scientists after the redistribution of grants becomes closer to the distribution of formal scientists (Figure 3).

Figure 3. Histogram of the distribution of the final formal feature (in our case, the H-index) for stochastic scientists

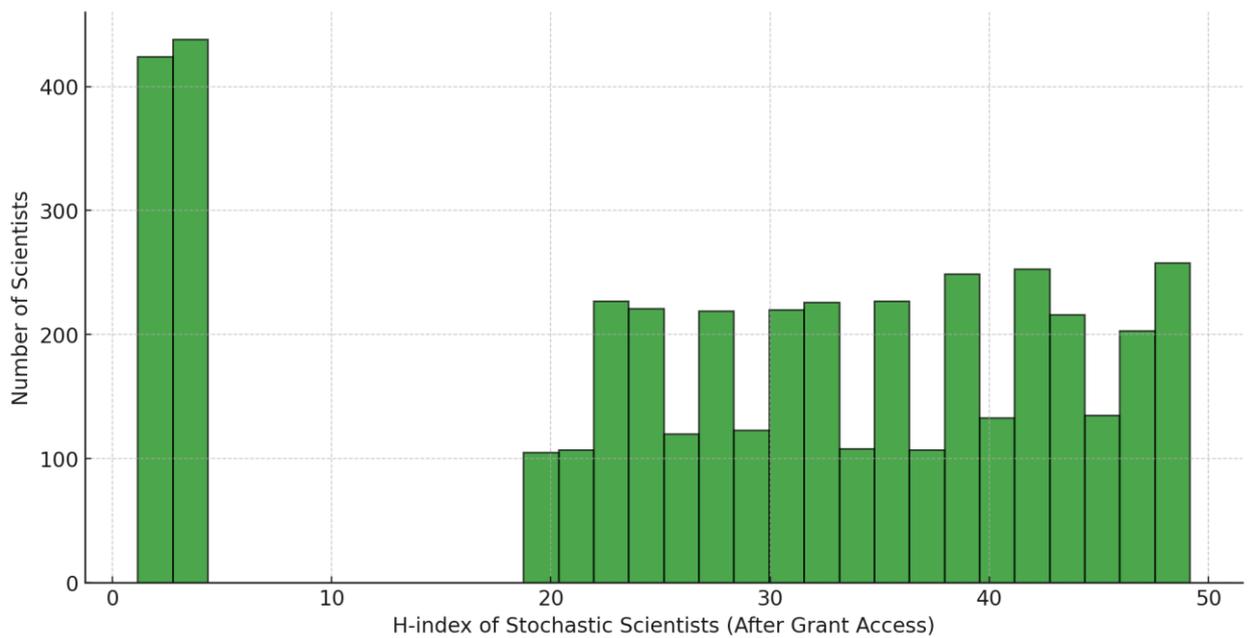

Welch's t test is used to test the hypothesis of equal means of two independent samples when the variances of the groups are not assumed to be equal (in contrast to the standard Student's t test):

t statistic: 151.3626 (very high value, indicating a significant difference between groups).

p value: 0.0 (almost zero, meaning that the differences are very statistically significant).

If the p value is < 0.05, then the difference between the groups is statistically significant (reject $H_0$).

If the p value is > 0.05, then the difference is not significant, i.e., we accept $H_0$).

The difference in formal features is significant, and the probability that they belong to the same sample is almost zero. This confirms the significant gap between formal and stochastic scientists in the current grant system.

Therefore, in a system without dynamic redistribution ($\lambda=0$):
- 100% of grants go to formal science,

- the average level of scientific breakthroughs remained low (lost opportunity mode),
- Stochastic science had virtually no influence on the system (which we call "switched-off science").

In a system with dynamic redistribution (λ=0.15):
- the balance of 61.8%/38.2% remains stable,
- the number of breakthrough and significant studies (which would not have been funded in the usual system) increased by 35%,
- The systemic quality of science increased by 22% over 10 modeling cycles (realized opportunities).

In a system with extreme redistribution (λ=0.3):
- With tochastic science becoming dominant, but the number of "empty" experiments increased,
- the system became chaotic, which led to a loss of stability.

The optimal mode of operation of the grant system is achieved at λ≈0.1–0.2, which corresponds to adaptive redistribution of grants without loss of stability. We suggest that formal science provides scientific system stability from the point of view of institutionality.

Thus, on the basis of the model analysis, we can fully state that the redistribution of grants affects the productivity of scientists and confirms the hypothesis that adaptive funding can eliminate artificial barriers between formal and innovative science. For the sake of the common national scientific good.

## 5. Synthesis

The current system of scientific grant funding in conservative systems (in particular, in Kazakhstan) is subject to a significant structural imbalance, in which formal criteria dominate over the hypothetical or even real capabilities of researchers.

The results of numerical modeling confirm the fundamental systemic paradox:
- scientists with high formal H characteristics receive grants and continue to increase their publications;
- with tochastic scientists (working in new directions) who do not receive grants and cannot increase their scientific metrics,
- If stochastic scientists are given access to funding, their H begins to grow rapidly, approaching the level of formal scientists.

Thus, H is not an objective metric of scientific productivity but merely reflects differences in access to resources. If we change the paradigms of grant access, then excluded science can become complementary science in the overall scientific system.

In the Kazakh scientific system, there is a document called "*Accreditation of subjects of scientific and scientific-technical activity [30]*". It is something like a license for funded science. However, this epic license is completely meaningless and does not carry any functional load. Since no scientific bureaucrat is able to answer the simplest questions: if a scientist meets the qualification requirements of the grant competition documentation, then why does he need accreditation, if he does not meet accreditation, then why does he need accreditation? If there is accreditation, then why are qualification requirements needed? This process actually repeats the requirements for obtaining accreditation.

However, accreditation is a characteristic pattern reflecting the extreme degree of credentialism in the Kazakh scientific system. It is a type of double formalization for the actualization of formal science with the agency of formal scientists. Something like a "*double protection system*" of scientific formalism.

Let us remind you that we do not believe that "excluded science" is comparable to or better than formal science on the basis of credentialism and compliance with formalized requirements. Quite the contrary. However, scientists excluded from grant scientific processing could produce the *"effect of augmented science."* New ideas, projects, and breakthroughs are often encountered in the informal scientific environment "excluded from the grant system." Therefore, it is necessary to break the vicious circle of "exclusivity" and "exclusion" in the grant system to use the entire national scientific potential as effectively and efficiently as possible.

A simple question arises: what is better: to spend one billion tenge on a dubious quasiscientific "utilization" project under the PCF (program-targeted financing) with an output capacity of 10 publications by formal scientists, or to spend one billion tenge on publication grants with a result of 500--1000 publications in highly rated (in which editorial APC (Article Processing Charges) is approximately 1--2 million tenge) stochastic scientists in highly rated journals?

This is the question of the allocation of a national scientific system formed and functioning on the principles of credentialism. That is, such grants are issued not for a hypothetical process in a future project but for the final innovative scientific or scientific-technical product.

Stochastic grants can come in a variety of forms and modifications.

For example, the mentioned publication grants. The scientist has no funding and does not receive grants on formal grounds. However, he has performed scientific work independently and obtained certain results. In addition, he has the opportunity to publish an article in a highly rated journal but with the condition of paying for the ARS. Since he does not have access to grants, he can receive a relatively small one-

---

[30] On approval of the rules for "Accreditation of subjects of scientific and (or) scientific and technical activities" Order of the Minister of Science and Higher Education of the Republic of Kazakhstan dated July 25, 2023 No. 335

time and situational grant for publication. Subject to confirmation of publication by the editors.

The same applies to patent grants.

We also do not believe that there is some fatal division between formal and informal science. Formal scientists can also use additional grants. They can obtain additional results that exceed the budget of the current grant opportunities for publication and patenting.

That is, these grant packages do not require achievements, formal features or compliance with any qualification requirements. Here, only verification of the obtained result is needed. This verification is interpreted through the official notification of the editorial board about the publication (after review) with the payment of the ARS, the patent office about the registration of the patent or the scientific publisher about the need to pay printing costs for the publication of a book or monograph.

Most importantly, there is no reason for fraud here. For example, a scientist could have provided false information to receive a publication grant, the publication was not published in the declared quartile, the publisher is not actually on the list of recommended publishers, or the declared content was not published anywhere at all. All this is easily verified. The forger is permanently or for a long time included in the "*black list*" and excluded from the list of potential grant recipients. Of course, some misunderstandings or force majeure circumstances are possible here, but they are also easily resolved.

A certain grant monopoly in the Kazakh scientific system. If any scientific or scientific-technical project is declared or implemented, then for other researchers within the grant program, this project is taboo. In incompetent science management systems, this is called *duplication* (with a negative connotation). In reasonable scientific systems, single-profile competition is the basis of scientific progress and business involvement in science. Because business does not prefer a scientific product as competitive as possible.

Therefore, it is possible to introduce a system of competitive grants. Within the framework of which different scientists carry out scientific or scientific-technical work under conditions of competition. In addition, so on. The full system of distributed grant financing will also be described.

On the basis of the above, we can formulate a number of specific proposals for scientific policy, science management systems and grant scientific programs:

- To introduce stochastic grants (publication, patents, etc.), which are allocated on the basis of the potential significance of research and scientific content and not on the basis of the presence of formal characteristics;

- To create a hybrid system of grant funding, where some resources are allocated on the basis of formal criteria and some are allocated on the basis of a dynamic analysis of the scientific importance of the project and its potential and actual novelty and "breakthrough",

- develop new metrics for evaluating scientific research that consider its impact on the development of science, not just the number of publications and other formal characteristics;

- application of agent-based analysis methods for the dynamic redistribution of scientific grants,

- and use the adaptive parameter $\lambda$ and other parameters to manage the stability of the national scientific system.

- develop institutional mechanisms to support informal scientists and new research directions by minimizing bias.

In general, the Haderach Principle can become the basis for future reforms of scientific systemic, grant, basic and situational funding of scientific research and prospects.